\documentclass[twocolumn]{aastex6}

\usepackage{natbib}
\usepackage{epsfig}

\newcommand{\sub}[1]{\ensuremath{_{\mbox{\scriptsize#1}}}}

\begin{document}

\slugcomment{To appear in ApJ}

\title{Gaps in Protoplanetary Disks as Signatures of Planets:
III.~Polarization}
\author{Hannah Jang-Condell}
\affiliation{Department of Physics \& Astronomy, University of Wyoming, 
  Laramie, WY 82071, U.S.A.}
\email{hjangcon@uwyo.edu}

\begin{abstract}
  Polarimetric observations of T Tauri and Herbig Ae/Be stars
  are a powerful way to image protoplanetary disks.  However,
  interpretation of these images is difficult because the
  degree of polarization is highly sensitive to the angle of
  scattering of stellar light off the disk surface.  
  We examine how disks with and without gaps created by planets 
  appear in scattered polarized light as a function of inclination
  angle.
  Isophotes of inclined disks without gaps are distorted in
  polarized light, giving the appearance that the disks
  are more eccentric or more highly inclined
  than they truly are.  Apparent gap locations are unaffected by
  polarization, but the gap contrast changes.
  In face-on disks with gaps, we find that the brightened
  far edge of the gap scatters less polarized light than the rest of the
  disk, resulting in slightly decreased contrast between
  the gap trough and the brightened far edge. 
  In inclined disks, gaps can take on the appearance of being localized
  ``holes'' in brightness rather than full axisymmetric structures.
  Photocenter offsets along the minor axis of the disk
  in both total intensity and polarized intensity images
  can be readily explained by the finite thickness of the disk.  
  Alone, polarized scattered light images of disks
  do not necessarily reveal intrinsic disk structure.
  However, when combined with total intensity images, the orientation
  of the disk can be deduced and much can be learned about disk
  structure and dust properties.  
\end{abstract}


\section{Introduction}

Polarized light imaging is being increasingly 
used as a method to obtain
high contrast images of the protoplanetary disks of young T Tauri and
Herbig Ae/Be stars.
This is because the light from the star is expected to be 
unpolarized, while light scattered from the disk should 
be preferentially polarized.
Polarimetric observations are therefore able to image disks to
small inner working angles because total intensity imaging requires much
more sophisticated PSF subtraction.  Techniques for starlight subtraction
in total intensity images such as angular differential imaging (ADI)
work well for point sources, but
end up self-subtracting axisymmetric features such as disks.  

Several facilities are producing high-quality images of
young circumstellar disks using the polarized differential imaging
technique.  
The Gemini Planet Imager (GPI) has demonstrated success with
imaging nearby bright debris disks in both total scattered light
and linearly polarized scattered light 
\citep[e.g.][]{2015Perrin+,2016Wolff+,2015Kalas+}.
The SPHERE (Spectro-Polarimetric High-contrast Exoplanet REsearch)
instrument on ESO's VLT (Very Large Telescope) has begun producing
high contrast images of exoplanets and disks in polarized light 
as well \citep{2016Garufi+,2016Olofsson+}. 
The Strategic Explorations of Exoplanets and Disks with Subaru (SEEDS)
survey team has produced a number of high contrast images of
young protoplanetary disks in T Tauri and Herbig Ae/Be objects
\citep[e.g.][]{2015Grady+,2015Akiyama+,2010Thalmann_etal,2011Hashimoto+,2012Mayama+,2013Follette+}.

Both debris disks and protoplanetary disks represent important but different 
stages of planet formation.  Protoplanetary disks,
often observationally categorized as T Tauri or Herbig Ae/Be stars,
are younger, optically thick,
and consist primarily of gas.  They represent the stage of
active giant planet formation, since these planets must form
while a large reservoir of gas still exists from which they can accrete
their massive envelopes.  Debris disks are older, optically thin,
and depleted in gas.  Because protoplanetary
disks are so young ($\lesssim10$ Myr), they are associated with
star-forming regions, which tend to be distant ($\gtrsim140$ pc)
from the Earth.
It is often easier to observe and image debris disks
because they occur in older stars and are not constrained to
star formation regions.

The focus of this paper is protoplanetary disks.  Because they are optically
thick, radiative transfer modeling of them is more complicated
than in optically thin debris disks.  As seen in the case of
HR 4796A \citep{2015Perrin+}, even a moderate optical depth 
can complicate the interpretation of observations.  
The most satisfactory explanation for near side/far side anisotropies
observed in the disk of HR 4796A that reconciles both the total intensity
and polarized intensity images is that there is a small amount of optical
depth that attenuates the emission from the near side of the disk.
This paper focuses on younger gas-rich optically-thick protoplanetary
disks rather than older gas-poor optically thin debris disks such as
HR 4796A.  However, it is an excellent 
example of confusion resulting from simplistic 
interpretations of polarized scattered light images

Stellar irradiation is the primary heat source in protoplanetary
disks. Thus, in order to accurately model the disk structure,
it is important to properly model the illumination of the disk surface
and carry out detailed radiative transfer calculations for the 
heating of the disk interior.  
Further radiative transfer calculations are needed to 
model observable properties of protoplanetary disks.
As in the case of HR 4796A, while polarimetric observations of
disks can produce very high contrast images, taken alone without
properly accounting for the disk orientation and optical depth, 
they can lead to misinterpretation of disk structure.  
Since the degree of polarization 
of scattered light depends on the angle of scattering,
the appearance of the disk is highly sensitive to 
the precise geometry of the disk, including the angle of inclination 
and features such as gaps.  Therefore, apparent structures seen in
scattered polarized light must be treated with caution. 

As a case in point, some of the first observations of AB Aurigae in
polarized scattered light showed what appeared to be a dark spot,
possibly caused by the presence of a planet \citep{Oppe08,2009HJC}.  
Follow-up efforts showed that the lack of emission was real, but
that it was not indicative of a local perturbation to the disk structure,
but rather caused by the inclination of the disk decreasing the
polarization fraction along the disk minor axis 
\citep{HJCKuchner,2009Perrin_etal}.
Further observations of AB Aurigae in scattered polarized light were
achieved at a small inner working angle,
revealing still more complex morphology in the inner disk
\citep{2011Hashimoto+}.
The apparent gaps, holes, and spiral arms were attributed to planet
formation, but little effort has been made to produce a model for
the disk morphology that accounts for the difference between
total and polarized scattered light intensity.  
Because such small inner working angles are difficult
to achieve with total intensity images, it is difficult to confirm
whether these features represent real structure in the disk.

In light of the capabilities of new instruments and the growing use of
polarimetry to detect and characterize disks, it is important to
fully understand how polarized light images differ from total intensity
images.  In this paper, we model scattered light imaging of
protoplanetary disks with and without gap clearing by growing planets.  
We examine how polarized intensity and total intensity image of the
same disks differ and how that varies with disk inclination.  

In previous work, we examined the thermal effects of gap creation 
by planets in protoplanetary disks 
\citep[][henceforth Paper I]{HJCTurner} 
and observable signatures of such planets 
\citep[][henceforth Paper II]{HJCTurner2}.
In this paper, we examine the appearance of the same set of disk 
models in polarized scattered light.  
We focus this study on the polarization of scattered light in the 
near-IR (1 micron), although these methods can apply to polarization 
by scattering at any wavelength, provided that a Rayleigh-like 
polarization law applies.
We do not consider polarization of thermal emission as might occur with 
grains aligned with magnetic fields which might be observed with 
ALMA\@.  

In \S\ref{sec:methods}, we describe the models used for 
calculating polarized scattered light images of disks with 
gaps created by embedded planets.  
In \S\ref{sec:results}, we show predicted images of protoplaneary disks 
in scattered polarized light, both with and without gaps.  
In \S\ref{sec:concl}, we present our conclusions.

\section{Methods}
\label{sec:methods}

\subsection{Disk Models}

The parameters for the disk models examined in this paper are described in
detail in Paper I \citep[see also][]{HJC_model,2009HJC}.
In particular, the
stellar mass is 1 $M_{\sun}$, the stellar radius is 2.6 $R_{\sun}$,
the stellar effective temperature is 4280 K,
consistent with a 1 Myr old protostar \cite{siess_etal}. 
The accretion rate of the disk is $10^{-8}$ $M_{\sun}$ yr$^{-1}$,
and the viscosity parameter is $\alpha=0.01$,
fairly typical of values for T Tauri stars.  The dust is
well-mixed with a composition consistent with
\citet{pollack_dust}, a maximum grain size of 1 mm,
a minimum grain size of 0.005 $\mu$m, and
a collisional grain size distribution, $N(a)\propto a^{-3.5}$
\citep{Dohnanyi1969}.

As described in Paper I, the disks are calculated with self-consistent
radiative transfer to determine the density and temperature structure
using the methods of \citet{HJC_model,2009HJC}.
The strategy used for calculating radiative transfer involves
spatially decomposing the disk surface into discrete planar surfaces,
each of which can be treated locally as plane-parallel.  Using an
analytic solution to the one-dimensional radiative transfer problem,
the amount of radiative heating at any given point in the disk is then
calculated as the summation of contributions to the heating from the
individual discrete surface elements.  

We impose gaps in the disk consistent with planets on circular orbits
at a semi-major axis of 10 AU and masses of 70 and 200 $M_{\earth}$,
as discussed in Paper I\@.
The adopted gap parameters are shown in Table \ref{fittable}
and are calculated based on 
viscous gap opening simulations by 
\citet{bate} and \citet{2006CridaMorbidelliMasset}.
The gaps are partially cleared,
in that the vertically integrated
surface density of the disk is decreased by no more than
60\% for the most massive planet modeled.
The gaps are modeled as Gaussian perturbations of the form
\begin{equation}\label{eq:gap}
  \label{gapdenprof}
  \Sigma(r) = \Sigma_0(r) \left\{1-d\exp[-(r-a)^2/(2w^2)]\right\}.
\end{equation}
where $\Sigma_0$ is the unperturbed surface density profile,
$d$ is the depth of the gap, $a$ is the gap position, and
$w$ is the gap width.  For a full discussion of the relationship
between planet mass, disk properties, and gap parameters,
see Paper I.  

\begin{deluxetable}{ccc}
  \tablecaption{\label{fittable}Gap parameters for given
    planet masses.}
\tablehead{
  \colhead{Planet mass\tablenotemark{1}} &
  \colhead{gap width ($w/a$)} & 
  \colhead{gap depth ($d$)}  
}
\startdata
72 $M_{\earth}$ (0.23 $M\sub{Jup}$) &
0.11 &
0.56 \\
210 $M_{\earth}$ (0.65 $M\sub{Jup}$) &
0.17 &
0.84\\
\enddata
\tablenotetext{1}{Actual masses used for this work.  In the text,
the masses have been rounded to 70 and 200 $M_{\oplus}$
for convenience.}
\end{deluxetable}

\subsection{Polarized Scattered Light}

As shown in Paper II, the scattered light brightness of the 
disk surface, assuming single isotropic scattering, is 
\begin{equation}\label{eq:sglscat}
I_1^s(\nu)=\frac{\omega_{\nu}\mu\/R_*^2\/B_{\nu}(T_*)}{4r^2(\mu+\cos \eta)},
\end{equation}
where $\mu$ is the cosine of the angle of incidence of stellar light on
the surface, $\omega_{\nu}$ is the wavelength-dependent albedo, $r$ is
the distance between the surface and the star, $\eta$ is the angle of
scattering to the observer with respect to the surface normal,
and $B_{\nu}(T_*)$ is the stellar
brightness at $\nu$, evaluated as the Planck function.

However, multiple scattering cannot be ignored, especially for 
grains with high albedo.  For photons scattered two 
or more times, the contribution to disk brightness is 
\begin{eqnarray}\label{eq:multscat}
I_2^s &=&   
\frac{B_{\nu}(T_*) R_*^2}{4 r^2}
  \frac{\mu \omega^2}{1-g^2\mu^2} \times \\
&&\quad  \left[ \frac{2+3\mu}{(1+2g/3)}
    \frac{1}{(1+g\cos\eta)}
      - \frac{3\mu}{(1+\cos\eta/\mu)}
\right]\nonumber
\end{eqnarray}
where $g=\sqrt{3(1-\omega_{\nu})}$ and isotropic scattering is assumed.  
The total scattered light intensity is then 
\begin{eqnarray}\label{eq:totscat}
I_{\nu}^s &=& I_1^s + I_2^s = 
\frac{\omega_{\nu}\mu\/R_*^2\/B_{\nu}(T_*)}{4r^2(\mu+\cos \eta)}
\times\\&&
\left\{
1 + \frac{\omega_{\nu}}{1-g^2\mu^2} 
  \left[\frac{(2+3\mu)(\mu+\cos\eta)}{(1+2g/3)(1+g\cos\eta)} - 3\mu^2
\right]
\right\}
\nonumber
\end{eqnarray}

To calculate the amount of linear polarization of scattered light, 
we make the assumption that photons scattered more than once become randomly
polarized, resulting in net zero polarization.  Thus, we need only
consider the polarization signal from the singly scattered photons.  
This reduces the polarization fraction overall, since 
multiple scattering contributes significantly 
to the disk brightness in total intensity scattered light,
as shown in above and in Paper II.  

The Rayleigh law for linear polarization 
states that the fractional polarization is given by
\begin{equation}
  \label{eq:Rayleigh}
f_R = \frac{P}{I} = \frac{(Q^2+U^2)^{1/2}}{I}
= \frac{\sin^2\theta}{1+\cos^2\theta} 
\end{equation}
where $\theta$ is the angle of deflection, 
and $I$, $Q$, and $U$ are the canonical Stokes parameters.
The Rayleigh regime is valid when the wavelength of light is
large compared to the size of the scattering particles,
or $\lambda\gg 2\pi a$.  
For real grains, the polarization phase function depends on
grain shape and composition in addition to grain size.  In
general, the shape of the polarization phase function is
qualitatively similar to the Rayleigh law.  For simplicity, we
assume a Rayleigh law for this work.  This is in part justified
by the fact that although 
the assumed dust composition
spans a large range of dust sizes, the scattering is dominated
by the smallest grains because their collective cross-section
is larger than that of the largest particles.

The polarization of singly scattered photons 
obeys the Rayleigh law.  If photons are multiply scattered, we assume that
their scattering angles are randomly reoriented, with the net effect being
random polarizations, so that the polarization of the ensemble of
multiply scattered photons is zero.
Then the polarized intensity or $P$ image is given by 
\begin{equation}
P = f_R(\theta) I_1^s
\end{equation}
where $I_1^s$ is the intensity of singly scattered photons, given 
in Eq.~(\ref{eq:sglscat}).  
The total fractional polarization is given by 
$P/I = f_R I_1^s / (I_1^s + I_2^s)$ 
where $I_2^s$ is the intensity of multiply scattered photons, given
in Eq.~(\ref{eq:multscat}), and 
\begin{eqnarray}\label{eq:polfrac}
  f &=& \frac{P}{I} = \frac{\sin^2\theta}{1+\cos^2\theta}
  \times\\&&
\left\{
1 + \frac{\omega}{1-g^2\mu^2} 
  \left[\frac{(2+3\mu)(\mu+\cos\eta)}{(1+2g/3)(1+g\cos\eta)} - 3\mu^2
\right]
\right\}^{-1}.
\nonumber
\end{eqnarray}

\section{Results}
\label{sec:results}

We apply the models for scattered polarized light to the disk models
presented in Papers I \& II\@: namely, a disk without planets, 
a disk with a 70 $M_{\oplus}$ planet at 10 AU, and a disk 
with a 200 $M_{\oplus}$ planet at 10 AU\@.  
The disk is assumed to be a steady-state passively accreting 
$\alpha$-disk model, orbiting a 1 $M_{\odot}$ star.  
We assume that the planet opens a gap in the disk, which we 
model as an ad hoc axisymmetric perturbation on the disk structure.  
For details of the disk models and analysis of their structure, 
see Papers I \& II\@.  

\begin{figure*}
  \plotone{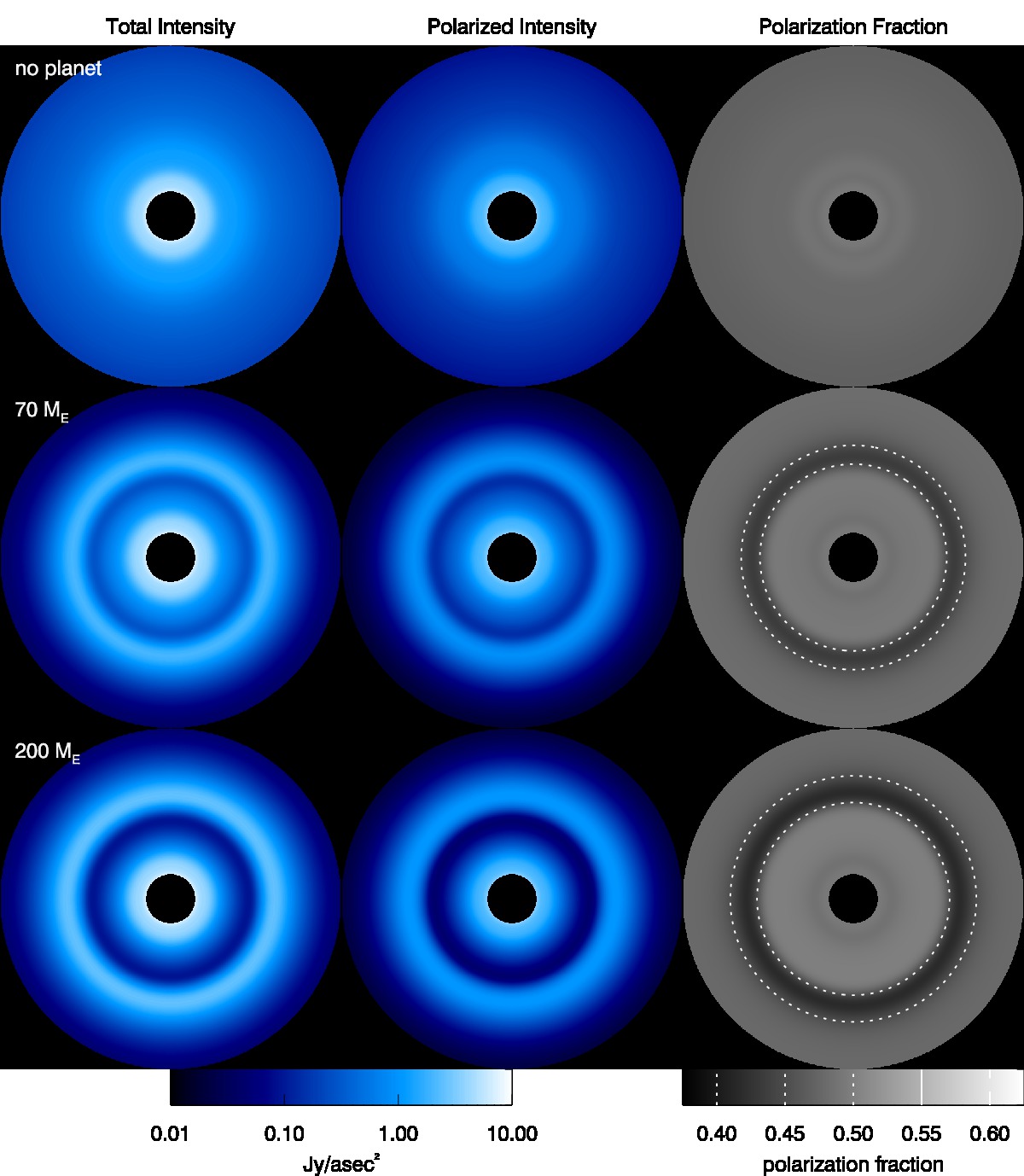}
  \caption{\label{fig:faceon}
    Model images of disks in face-on orientation ($0\degr$ inclination).
    Left: the total 1 $\mu$m scattered brightness from the disk ($I$).  
    Center: the $P$ image, or total polarized intensity. 
    Right: the polarization fraction, $f=P/I$. 
    The top, middle, and bottom rows show disks with gaps created
    by planets of mass 0, 70 and 200 M$_{\earth}$, respectively.
    The colors of the left two columns are on the same scale on the
    left in Jy/asec$^2$, while the color scale for the
    polarization fraction is on the right.  The dotted contour shows
    a polarization fraction of 0.4, and is identical in location
    between the right two columns.  
  }
\end{figure*}

We calculate the brightness of total scattered light and 
polarized light at 1 $\mu$m from the surface of the three disk models
described.  
Simulated images generated at J, H, and K bands
produce qualitatively identical results, with slight differences
arising from the variation of extinction coefficients and
albedos across wavelengths.  Compositional and grain
size variations can also result in variations in extinction and
albedo, which suggest that the overall results presented here
apply to a wide range of grain properties as well.
The opacities used to generate the models here
correspond to a well-mixed \citet{pollack_dust} dust composition with a
\citet{Dohnanyi1969} collisional size distribution with
minimum/maximum grain size of $a\sub{max}=1\,\mbox{mm}$
and gas-to-dust ratio of 0.138 
(see also \citet{2009HJC}).  For the record,
the extinction opacities at 1 $\mu$m is 
10.86 cm$^2$ per gram of gas, with an albedo of 0.9099.  

\subsection{Face-on Disks in Polarized Scattered Light}
\label{sec:faceon}

\begin{figure}
  \centerline{\includegraphics[width=3.3in,trim=0in 0in 0.2in 0.25in]{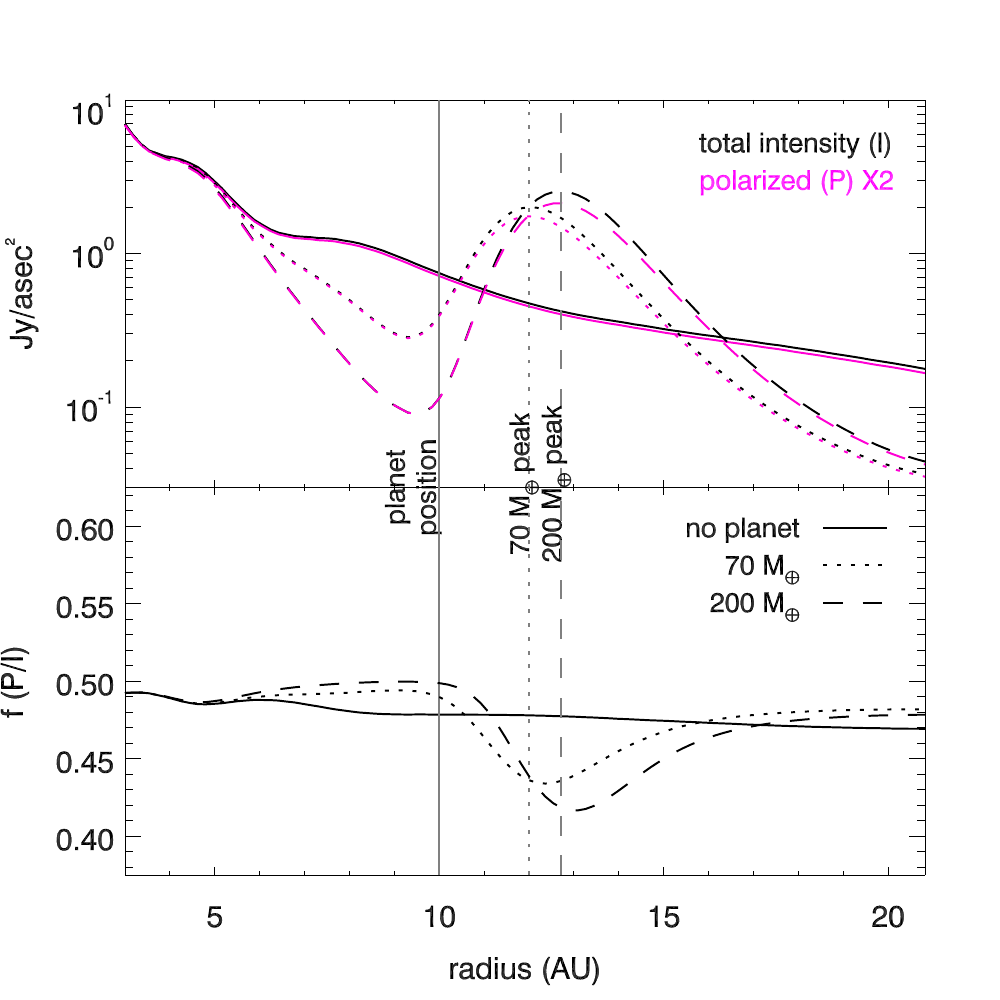}}
  \caption{\label{fig:profiles}
    Brightness profiles in face-on disk models ($0\degr$ inclination).
    Solid, dotted, and dashed lines indicate models with gaps created
    by planets of mass 0, 70 and 200 M$_{\earth}$, respectively.  
    Upper plot: black lines indicted total intensity ($I$) while
    magenta lines show polarized intensity ($P$)
    scaled by a factor of 2 to better compare to total intensity.
    Lower plot: polarization fraction ($f=P/I$).
    The position of the planet is indicated by the solid gray line.
    The dotted and dashed lines indicate the peak brightness in the
    total intensity images of the 
    disk models with 70 and 200 M$_{\earth}$ planets, respectively,
    showing that the peak brightness is coincident with a
    minimum in polarization fraction.  
  }
\end{figure}

In Figure \ref{fig:faceon} we show simulated images at 1 micron
of our fiducial disk model, with and without planets, at an
inclination of 0\degr (face-on orientation).
Three sets of images are shown for each disk: total scattered
intensity ($I$), polarized intensity ($P$), and
polarization fraction ($f=P/I$). 
The brightness profiles are also plotted in Figure
\ref{fig:profiles} to show greater detail.  

The polarization fraction is nearly constant across the face of
the disk without any gap, with a slight downward trend toward increasing
radial distance.  This is to be expected for a typical flared
protoplantary disk structure.  If the scattering surface of the disk
($z_s$) were flat so that $z_s/r$ were constant, then the polarization
fraction would be exactly constant over the face of the disk.
Because the disk is flared so that $z_s/r$ is steadily increasing with
$r$, the scattering angle increases and so the Rayleigh law
gives smaller polarization fraction.

Both the total intensity and polarized intensity images clearly show the
gaps for both the 70 and 200 $M_{\earth}$,
with a dark shadow in the gap trough and a brightened
outer gap edge. 
Considering only the polarized fraction, there is a region of
decreased polarization in an annulus centered not on the planet position
or gap trough but rather the brightened outer gap edge.
This can be explained not from the differences in scattering angle, but
rather in the proportion of total intensity from singly- versus
multiply-scattered photons.
In essence, while the viewed intensity of both singly- and
multiply-scattered photons increases with the cosine of
the angle of incidence at the disk surface ($\nu$), the intensity of
multiply-scattered photons increases at a faster rate.  
On the exposed outer gap edge, the angle of incidence is closer
to 0, so $\mu$ is larger, the total brightness increases, and
the fractional polarization goes down.  
This can be seen be examining equations
(\ref{eq:sglscat})-(\ref{eq:totscat}).
From Equation (\ref{eq:sglscat}), we see that
$I_1^s \propto \mu/(\mu+cos\eta)$. 
On the other hand, $I_2^s \propto \mu$ to first order in $\mu$ 
for $g\cos\eta\ll 1$.
The albedo of the assumed grain composition is is $\sim0.9$, so
this is a reasonable assumption.  For a thin disk, $\mu\ll1$ and
$\cos\eta\sim1$.  When the disk surface tilts up, the angle sum
$\eta+\cos^{-1}\mu$ stays roughly the same.  However, $\mu$ increases
approximately linearly while $\cos\eta$ decreases more slowly.
Thus, while both $I_1^s$ and $I_2^s$ both increase as $\mu$ increases,
$I_1^s$ increases more slowly.  

\begin{figure*}[p]
  \plotone{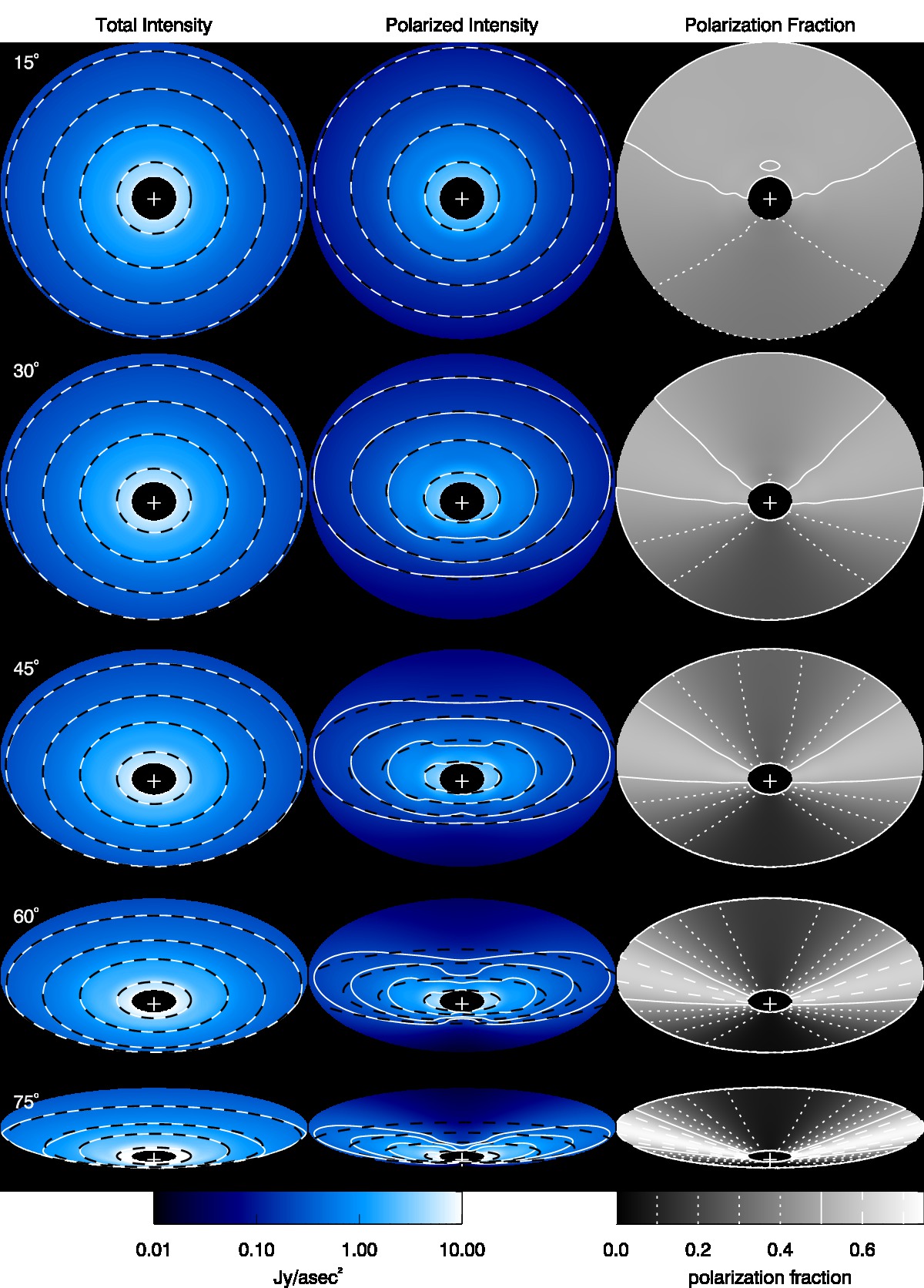}
  \caption{\label{fig:nogap}
    Model images of a gap-less disk observed in scattered total and
    polarized light at 1 $\mu$m at varying inclinations.
    Columns are arranged the same as in Figure \ref{fig:faceon}.
    From top to bottom:
    $15\degr$, $30\degr$, $45\degr$, $60\degr$, and $75\degr$
    inclination.
    The disks are tilted so that the top side of the image
    is the far side of the disk. 
    In the $I$ and $P$ images, white contours show isophotes 
    and black dashed lines show best-fit ellipses to those isophotes.
    Isophotes are chosen to have maximum radius of 5, 10, 15, and 20 AU\@.
    In the $f$ images,
    solid, dotted, and dashed contours show $f$ values of
    0.5, $<0.5$, and $>0.5$, respectively.  The values of the
    contours are indicated on the color scale bar.}
\end{figure*}

\begin{figure*}[p]
  \plotone{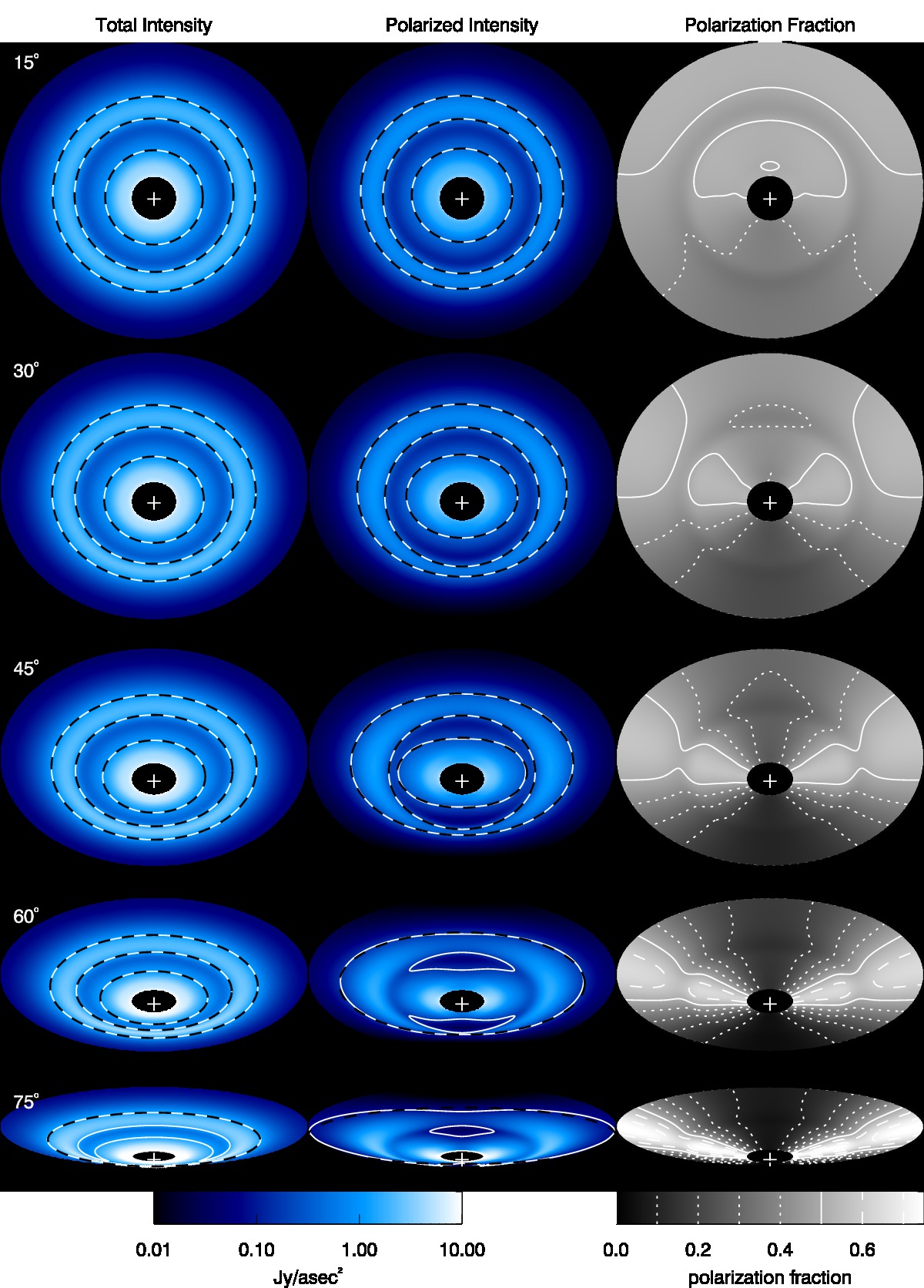}
  \caption{\label{fig:subcr}
    Same as Figure \ref{fig:nogap}, for a disk with a partial gap cleared 
    by a 70 $M_{\earth}$ planet.
    The isophotes here trace the brightness
    level equal to half the brightness of the brightened gap edge
    along the minor axis toward the far edge of the disk.  The two inner contours
    bracket the gap trough, while the two other ones bracket the brighten
    far edge of the gap.
}
\end{figure*}

\begin{figure*}
\plotone{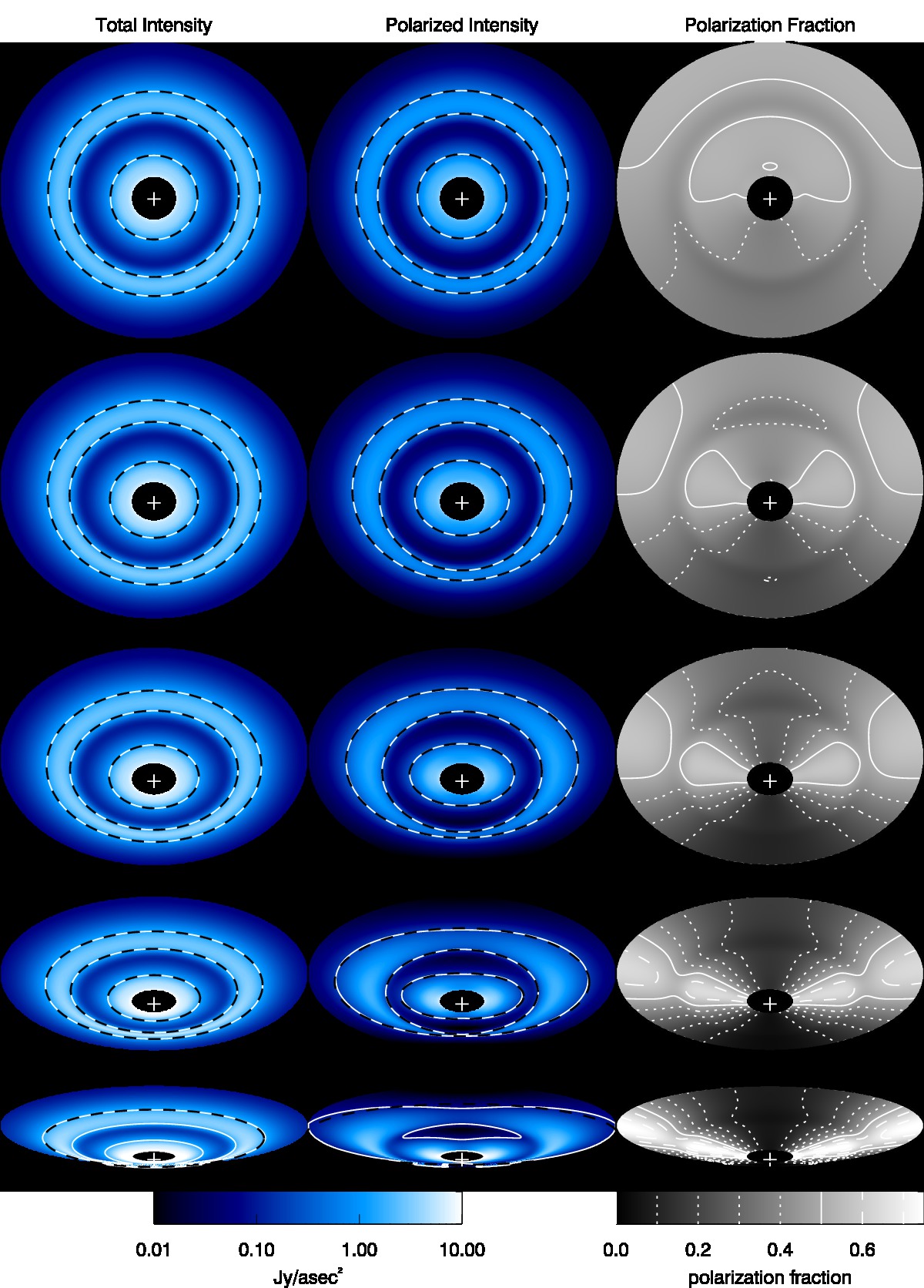}
\caption{\label{fig:mcrit}
Same as Figure \ref{fig:subcr}, for a disk with a partial gap cleared 
by a 200 $M_{\earth}$ planet.  
}
\end{figure*}

Specifically, $\mu$ is greater on the brightened
outer gap edge than elsewhere in the disk, meaning that
a greater proportion of scattered light comes from multiply-scattered
photons as compared to elsewhere in the disk.  Similarly, there
is a slight increase in polarization fraction within the
gap trough.  This results in a lowered gap contrast in scattered
polarized light imaging.  
The gap contrast, expressed as the ratio of 
maximum brightness of the outer gap edge to the 
minimum brightness in the gap trough, is 7.0 and 28
for the 70 and 200 $M_{\earth}$ models, respectively, in
the total intensity images.  
These values are reduced to 6.2 (12\% reduction) and
23 (16\% redection), respectively,
in polarized scattered light.  

This can also be expressed in terms of the fractional polarization
(Figure \ref{fig:profiles}, lower plot).
The fractional polariation stays relatively constant in the
disk with no planet, with a slight decrease toward larger
radii.  In the disks with gaps, there is a slight increase
in fractional polarization at the trough of the gap and
a decrease from about $48\%$ to $43\%$ ($41\%$) for the 70
(200) $M_{\oplus}$ case.  GPI has demonstrated the ability to
measure fractional polarizations to within a few percent
\citep{2015Perrin+}, so this subtle effect may be measureable.  

These results suggest that while the locations of gaps in
face-on disks can be determined from either total intensity
or polarized intensity images, the gap contrast might be
slightly underestimated from polarized intensity images alone.
On the other hand, if total intensity
images are also available, dark and bright lanes in disks
caused by shadowing and illumination of exposed surfaces
may be confirmed by the increased/reduction in polarization fraction
in shadowed/illuminated regions.

\subsection{Inclined Disks in Polarized Scattered Light}
\label{sec:tilted}

In Figures \ref{fig:nogap}, \ref{fig:subcr}, and \ref{fig:mcrit}, 
we demonstrate the effect of inclination on the appearance of disks 
in 1 $\mu$m scattered light.  
Images for a gapless disk are shown in Figure \ref{fig:nogap}, 
and disks with gaps created by 70 and 200 $M_{\earth}$ 
planets are shown in Figures \ref{fig:subcr} and \ref{fig:mcrit}, 
respectively.  
Each figure shows the predicted appearance of the disk 
with changing inclination angle 
when observed in total intensity and polarized scattered light.  
For reference, we also show the polarization fraction 
in the right column.
The morphology of the $P$ image in each realization
is significantly different from 
the $I$ image because of the spatial 
variation in polarization fraction.

\begin{figure*}[t!]
  \plotone{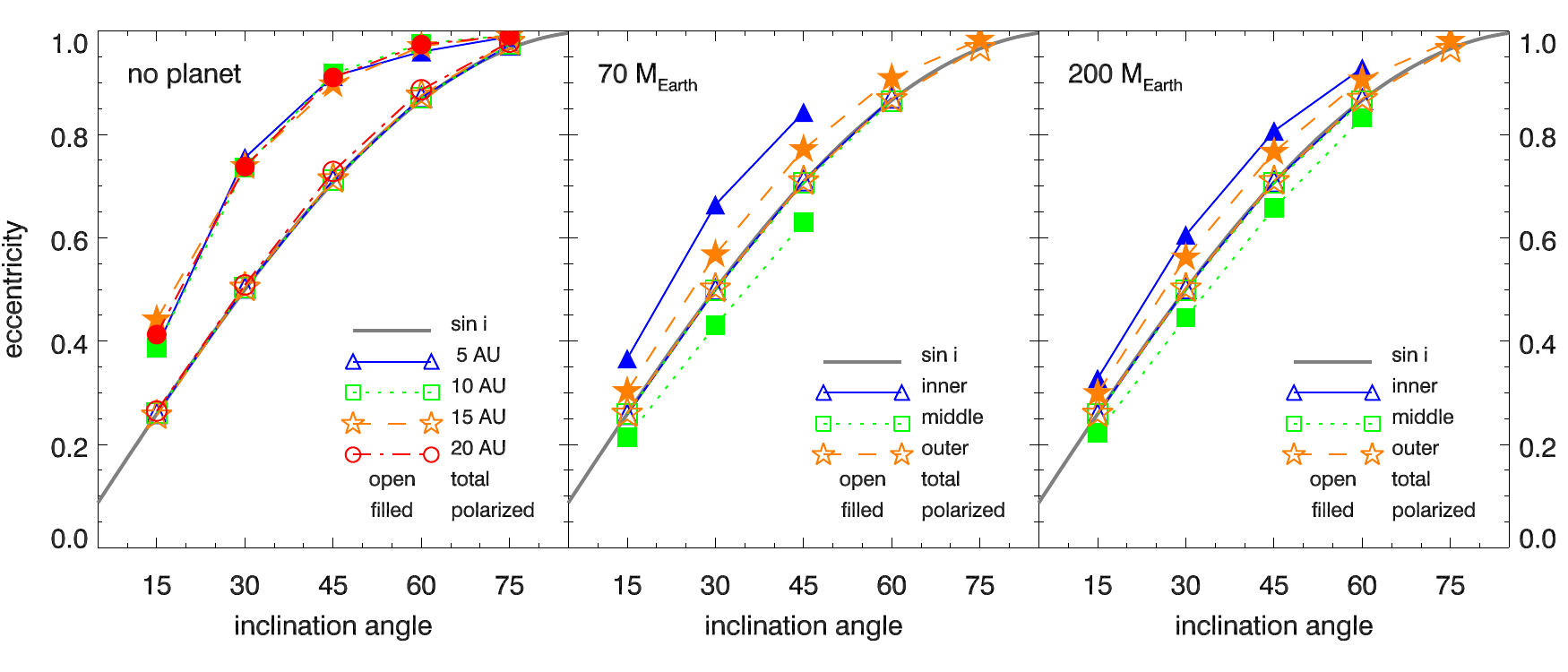}
  \caption{\label{fig:isoeccen}
    Eccentricity of elliptical fits to the isophotes shown in
    Figures \ref{fig:nogap}-\ref{fig:mcrit}
    as a function of disk inclination.
    For a flat radially symmetric disk, the expected eccentricity
    is $\sin i$, shown as a gray line.
    Note that eccentricities are systematically higher in the
    polarized intensity images.
    }
\end{figure*}

\begin{figure}
  \centerline{\includegraphics[width=3.35in]{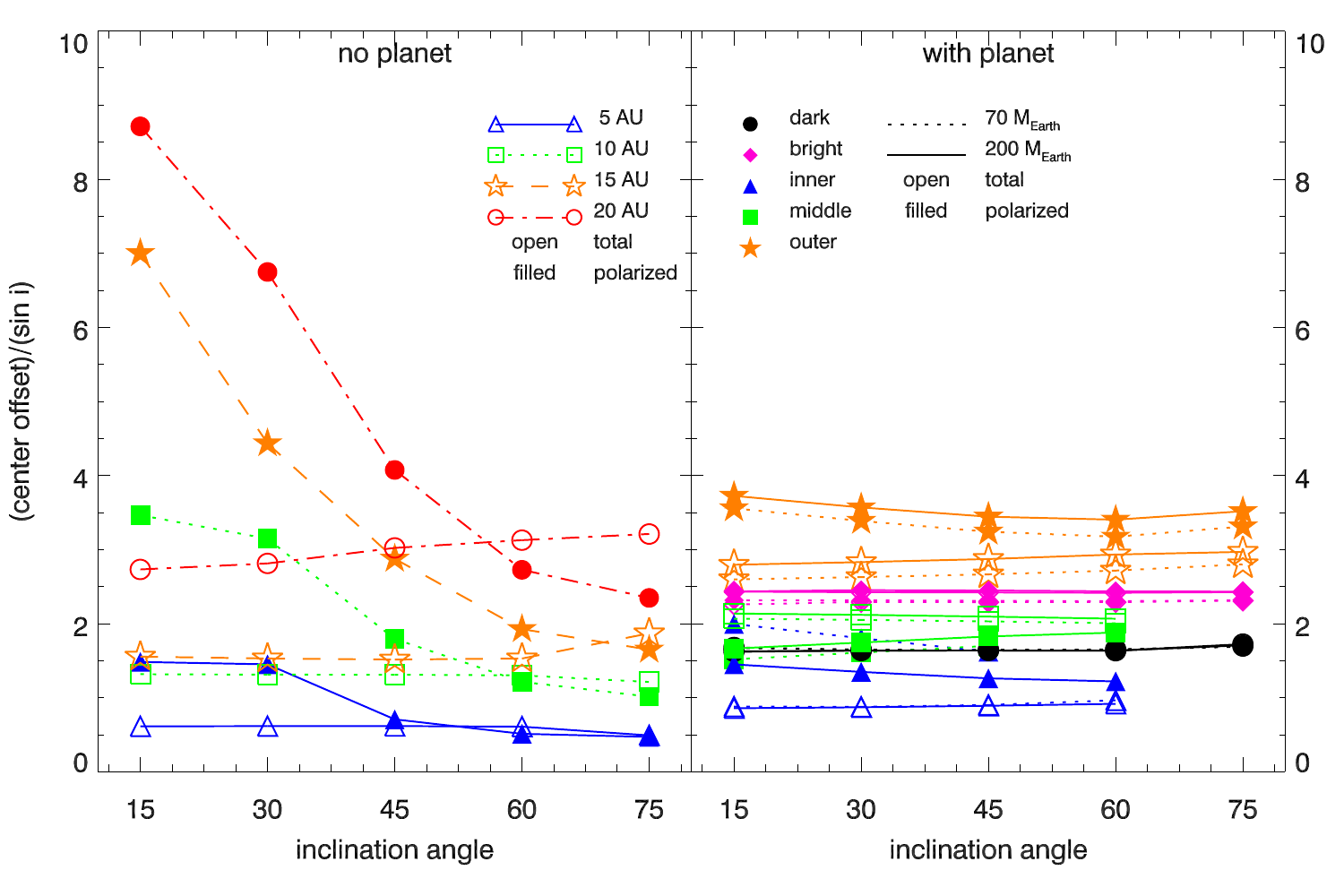}}
  \caption{\label{fig:isooffset}
    Offsets between stellar position and the center of the black dashed 
    ellipse fits in Figures \ref{fig:nogap}-\ref{fig:mcrit}
    as a function of inclination angle.
    Offsets are plotted for the planetless
    disk on the left, and gapped disks on the right,
    and are scaled by $\sin i$.
    If the offset were due to disk thickness alone, the
    lines should be strictly horizontal.
    Offsets for the dark gap trough and bright gap edge
    of 70 and 200
    $M_{\earth}$ planets are virtually identical between total and
    polarized intensity, so those points are nearly
    coindent.  
  }
\end{figure}

Figure \ref{fig:nogap} shows simulated images of the disk
with no gap.
In the total intensity images (left column), the isophotes
(white solid lines)
are well-fitted by ellipses
(black dashed lines)
with eccentricity equal to $\sin i$, which is the
apparent eccentricity of a circle inclined by angle $i$,
as shown in Figure \ref{fig:isoeccen}.  
However, the centers of the elliptical fits to the isophotes (photocenters)
are offset from the stellar position,
as shown in Figure \ref{fig:isooffset} (left panel).
This is due both to the
thickness of the disk as well as due to anisotropies in brightness
in scattered light from the near and far sides of the disks,
an effect discussed in detail in Paper II\@.

\begin{figure}
  \plotone{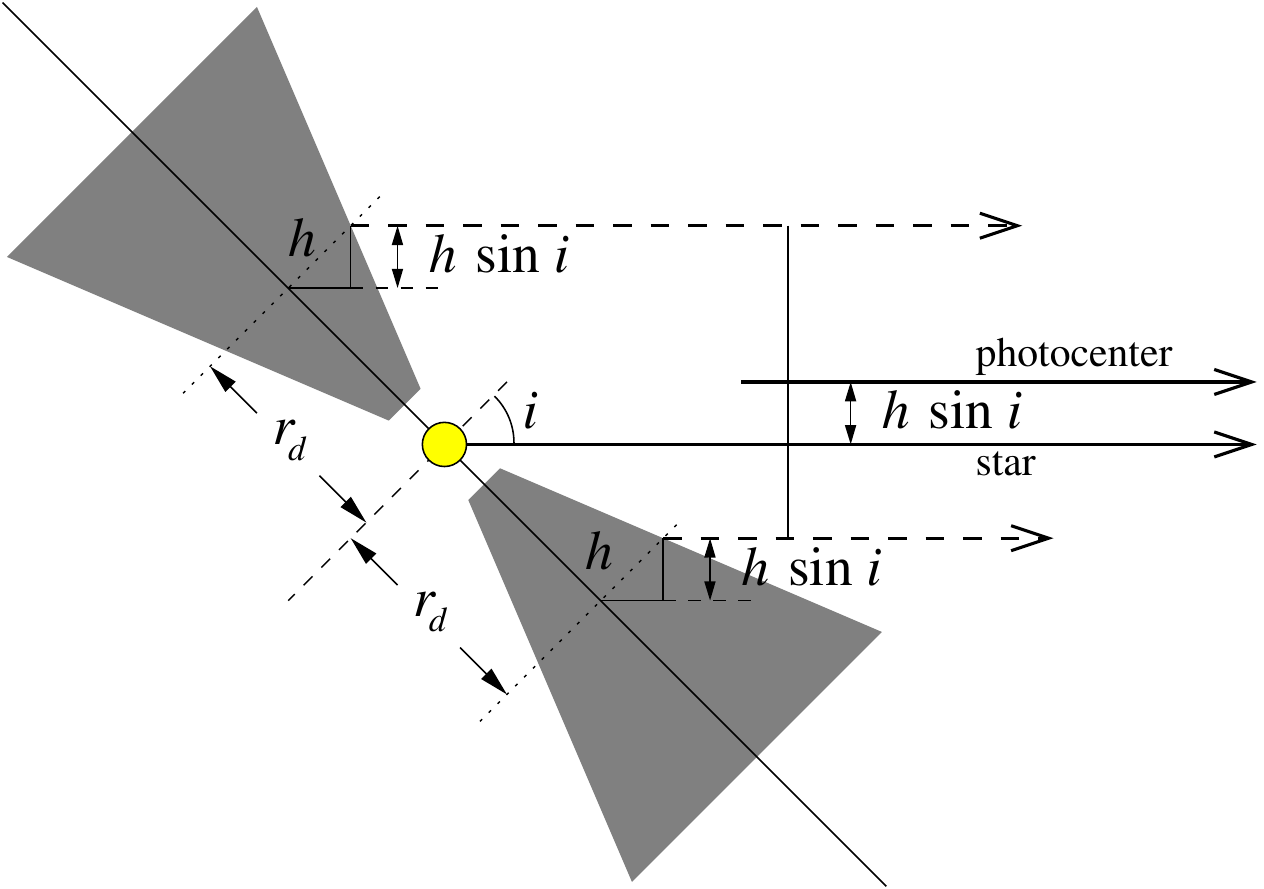}
  \caption{\label{fig:offsets}
    Illustration of how disk thickness leads to photocenter offsets.
    Points equidistant from the star at the disk surface appear offset
    from the stellar position by $h\sin i$ where $h$ is the disk thickness
    and $i$ is the inclination.  If an isophote traces azimuthally
    symmetric positions at the disk surface, then the photocenter will
    be offset from the stellar position by $h\sin i$.
    Note that whether the disk is flared or flat has no bearing on this
    effect: it is purely due to the finite thickness of the disk.  
  }
\end{figure}

Figure \ref{fig:offsets} demonstrates how the thickness of an inclined
disk creates a photocenter offset.  The fact that the disk scattering
surface is above the midplane of the disk means that an isotropically
scattering surface should have a photocenter offset of $h\sin i$
where $h$ is the height of the scattering surface above the midplane
and $i$ is the inclination.  This effect is modified somewhat because
the near side should be somewhat brighter than the far side, as
shown in Paper II\@. In Figure \ref{fig:isooffset}, the offsets are
scaled by $\sin i$ so that if the photocenter offset is dominated
by the geometric effect, the lines plotted on it should be horizontal,
with a value equal to the height of the scattering surface.  
This appears to be the case for the total intensity images
in the gapless disk (Figure \ref{fig:isooffset} left, open points),
with the height of the scattering surface increasing with distance
in the disk, as expected for a flared disk structure.  
  
The near and far sides of the disk appear dimmer in the $P$ images
in Figure \ref{fig:nogap} (center column), 
with the amount of dimming increasing with inclination.
This is as expected, since the scattering angles from 
the near and far sides of the disk are furthest from $90\degr.$
As shown in the plots of fractional polarization (right column),
the fractional polarization is smallest along the minor axis of the
disk because the near and far sides of the disk are most inclined
with respect to the observer.  
Thus, the disk appears more elliptical in $P$ images
versus $I$ images, as illustrated
from the shapes of the isophotes. 
The eccentricity of these isophotes, as calculated from the best-fit
ellipses, are plotted in Figure \ref{fig:isoeccen}.
As inclination increases, the isophotes deviate from ellipical shapes, 
tracing apparent ``divots'' in the $P$ image of the disk,
caused purely by the inclination geometry of the disk. 
The implication is that
disk inclination should not be inferred from $P$ images alone,
unless the disk structure is sufficiently well-understood to
deduce the polarization fraction.  

\begin{figure}
  \plotone{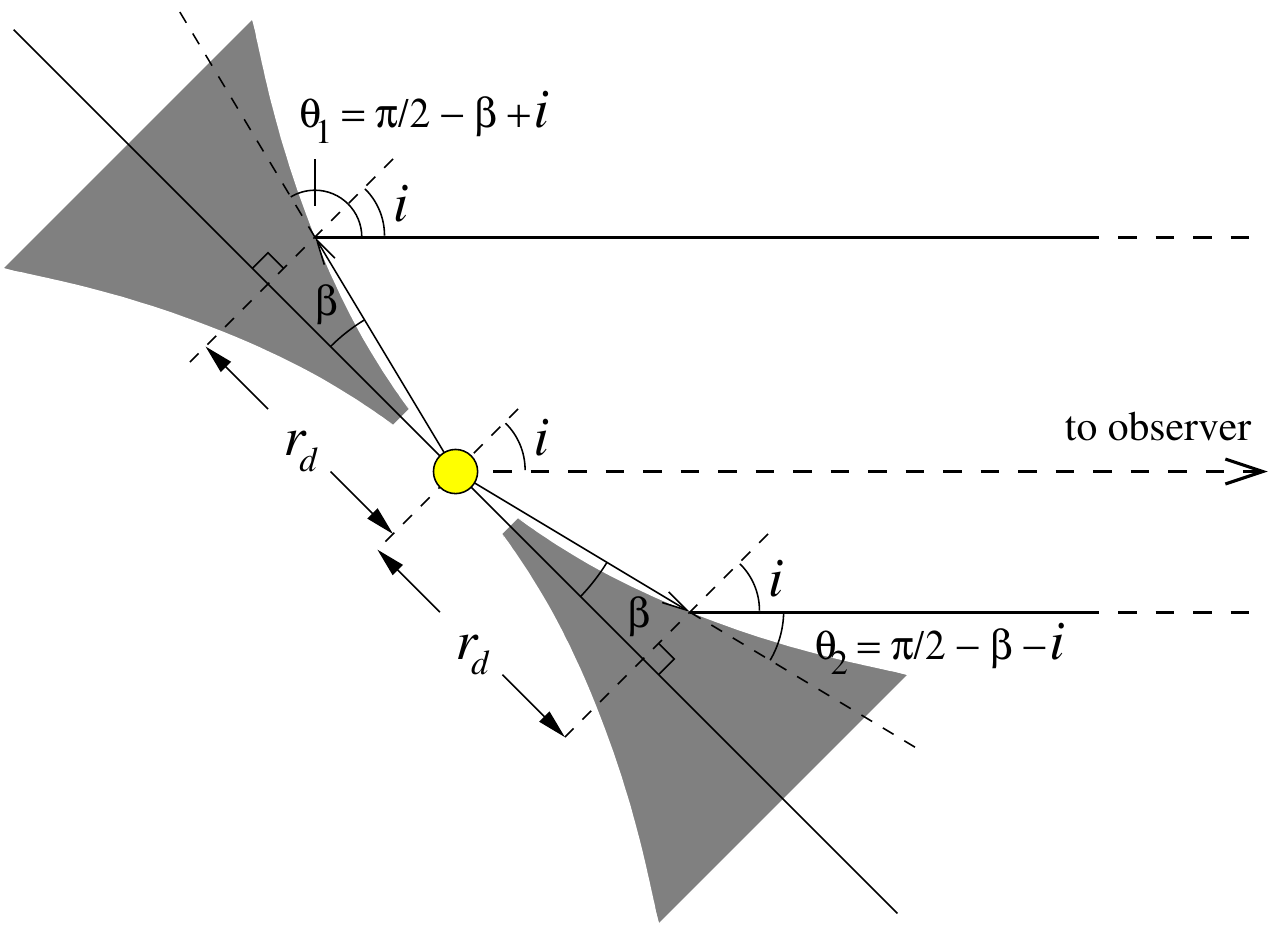}
  \caption{\label{fig:angles}
    Illustration of 
    scattering angles from opposite side of an inclined disk.
    The disk is inclined at an angle $i$ and the opening angle
    of the disk is $\beta$ at a distance $r_d$ from the star.
    Light is scattered off the disk surface toward the
    observer in a non-symmetric fashion because of the
    thickness of the disk.  Specifically,
    $|\theta_1-\pi/2|<|\theta_2-\pi/2|$.
    Since the peak in polarization fraction occurs at $\theta=\pi/2$
    and falls off away from that value, this means that
    the polarization fraction is higher on the far side of the disk.  
  }
\end{figure}

The polarization fraction is lower on the near side than the far side
due to the geometry of scattering off the surface of a disk with
finite thickness.  
This is because the scattering angle is farther from $90\degr$ 
on the near side than the far side, as illustrated in
Figure \ref{fig:angles}.
If the scattering angles off the far side ($\theta_1$) and
near side ($\theta_2$) were supplementary so that
$\theta_1+\theta_2=180\degr$, then the fractional polarization
as predicted from the Rayleigh law in Equation (\ref{eq:Rayleigh})
should be the same on both sides.
This would be true if the opening angle of the disk ($\beta$)
were zero, in which case $\theta_1 = 90\degr+i$ and
$\theta_2 = 90\degr+i$.  
However, because
the disk has a finite height, $\beta>0$, 
and $\theta_1$ and $\theta_2$ are no longer supplementary.
In particular, $\theta_1$ is closer to $90\degr$ than $\theta_2$,
so the fractional polarization is lower
on the near side of the disk.  

At high inclinations (60\degr and 75\degr),
the near and far sides of the disk
dim so much that apparent ansae form
in the
$P$ images of the disk.  This phenomenon is different from the ansae
often seen in images of optically thin debris disks.  In the debris
disks, the ansae are caused by an increase in column density
of scattering grains along the lines of sight through the disk along the
apparent major axis.  The disks modeled here are optically thick,
so changes in column density through the disk are unimportant.
Rather, the light scatters off the disk surface, so the brightness
depends on the angle of illumination at the disk surface and the
scattering angles along with the optical properties of the dust itself.  

In addition to distortion to the shapes of the isophotes,
at low inclination angles the photocenter offset is higher 
in polarized intensity 
as compared to total intensity, as seen in Figure \ref{fig:isooffset}
(left panel),  
While the photocenter offsets scale roughly with $\sin i$ in
total intensity (open points),
this is not at all the case in polarized light (filled points).
The reason for this is because of the anisotropy in polarization
fraction.
Since the near side of the disk has a smaller polarization fraction
than the far side, this has the effect of shifting the photocenter
toward the far side of the disk, increasing the amount of offset.
At high inclination angles, the effect becomes muted because the disk's
near edge becomes foreshortened as it tips up toward the observer.  

\begin{figure}
  \centerline{\includegraphics[width=3.5in,trim=.0in 0.1in 0in 0.4in]{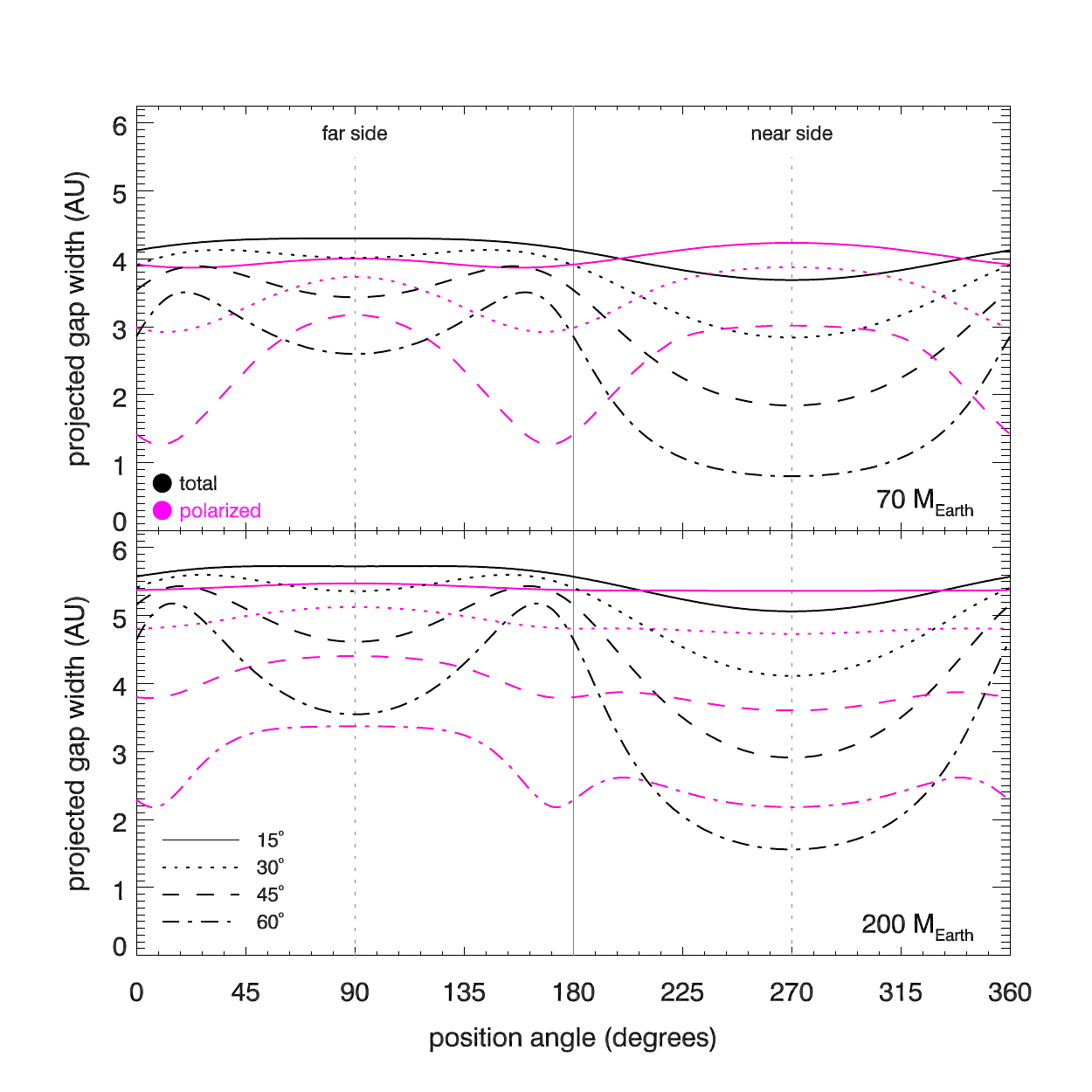}}
  \caption{\label{fig:gapwidth}
    Width of the gap as measured by the inner two isophotes traced in
    Figure \ref{fig:gappics}.  Top/bottom plots show 70/200 $M_{oplus}$
    planets.  Black/magenta lines trace the
    total/polarized intensity images, and line type indicates inclination
    angle.  Note that the apparent gap widths different greatly depending
    on whether they are observed in total versus polarized intensity.  
  }
\end{figure}

Figures \ref{fig:subcr} and \ref{fig:mcrit} show the effect of a gap
caused by a planet on the disk images.  To a large extent, the images
are a superposition of the tilted disk images seen in Figure
\ref{fig:nogap} with the decreased polarization fraction on the
brightened outer gap edge seen in Figure \ref{fig:faceon}.
Because the polarized intensity images appear darker along the minor axis
of the disk than in total intensity, this leads to 
an interesting effect where the width of the apparent gap
varies azimuthally in the $P$ images.
This is evident in the shapes of the isophotes overplotted in the
images in Figures \ref{fig:subcr} and \ref{fig:mcrit}.  The
isophotes in each total intensity and polarized
intensity image are set to half the brightness of the brightened
gap wall on the far side of the disk (upper part of the image).  
Whereas the apparent
gap width is narrowest along the image minor axis in the $I$ images,
in the $P$ images the opposite is true.
The gap widths as measured by the distance between the isophotes plotted
in Figures \ref{fig:subcr} and \ref{fig:mcrit} are shown in
Figure \ref{fig:gapwidth} to explicitly show this relation.
The gap widths in total intensity images
peak close to the major axis of the disk 
(position angles 0\degr and 180\degr), but in the polarized intensity
images they peak at the minor axis of the disk 
(position angles 90\degr and 270\degr).

This causes the inner disk
to appear to be ``warped'' in comparison to the outer disk, since the
ellipticity of the inner edge of the gap is higher than the outer edge
in the $P$ images.  If viewed with low angular resolution, the darkest
portions of the gap in the $P$ image
might appear to be dark ``spots'' rather than
part of a full ring.
We plot the eccentricities of the isophotes of the gapped disk images 
in Figure \ref{fig:isoeccen}
for the gapped disks in the middle and right panels.
In total intensity images, the eccentricities closely follow
$\sin i$, as expected for a simple inclined disk.  However,
the innermost isophote, tracing the inner edge of the gap shadow, 
is consistently more ecccentric than $\sin i$, while the
middle isophote, tracing the outer edge of the gap shadow, is 
consistently less eccentric than $\sin i$.  A naive interpretation
of the polarized images would lead to the conclusion that the inner disk is
tilted with respect to the outer disk, suggesting a ``warp'' in the
disk where none truly exists.
We also note that the outermost isophote is also consistently more
eccentric than $\sin i$, showing the difficulty of extracting disk
inclination solely from polarized intensity images.

The offsets of the isophote centers from the stellar position
as measured from the elliptical fits are plotted in
Figure \ref{fig:isooffset} (right).  As compared to the isophote
offsets in the disks without planets, the center offsets are
more well-behaved in disks with gaps in both total and polarized
intensity images, as if the gaps help to anchor the isophote position.
They scale roughly with $\sin i$, with the scale factor
increasing from the inner to the outer isophote, as expected from the 
$h\sin i$ relation (see caption of Figure \ref{fig:isooffset}).

\begin{figure*}
  \centerline{\includegraphics[width=7in]{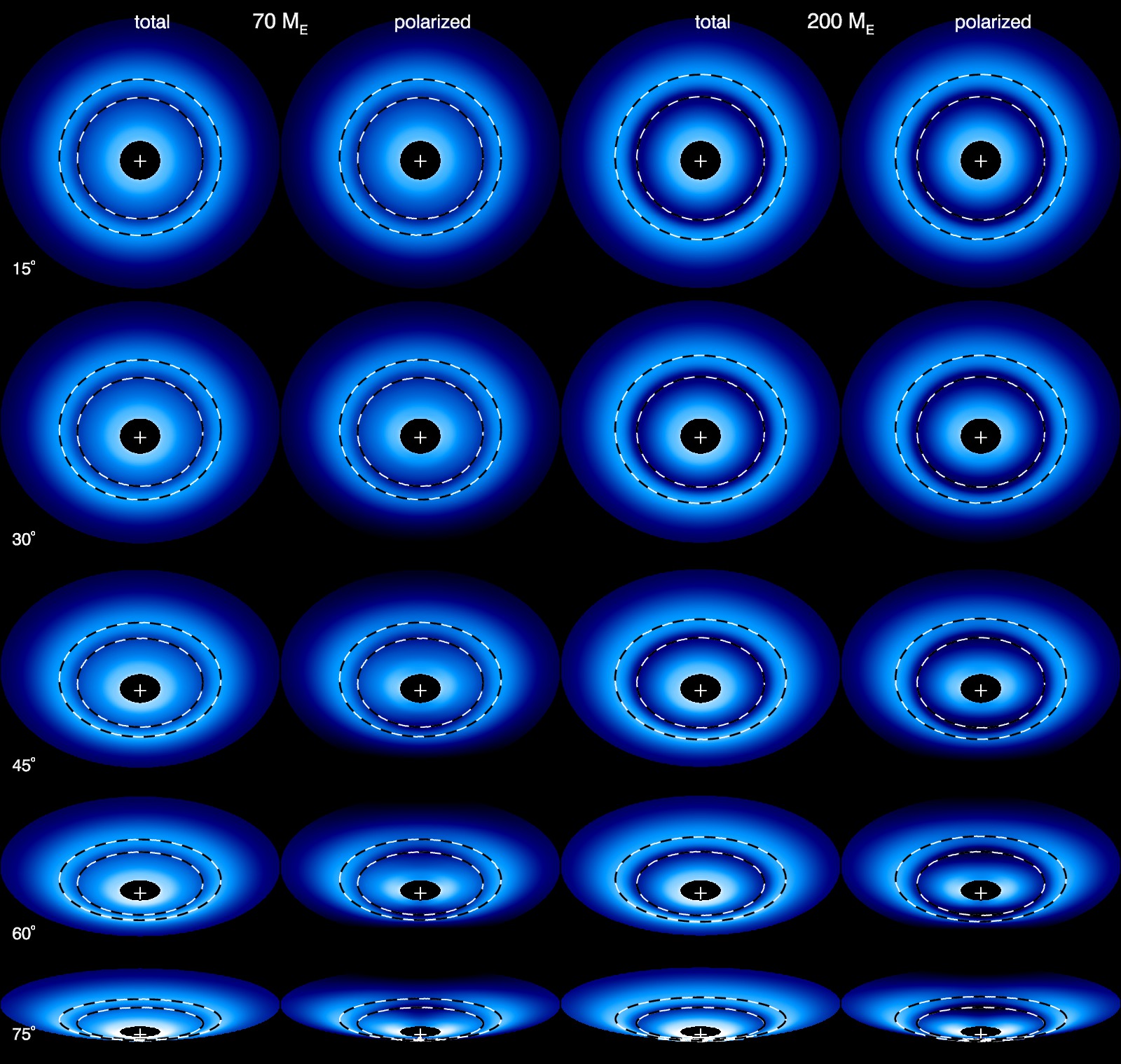}}
  \caption{\label{fig:gappics}
    Gapped disk images shown in Figures \ref{fig:subcr} and \ref{fig:mcrit},
    showing the gap positions.  
    The white lines show the radial brightness minima and maxima in
    order to trace the shadowing and brightening from the gap.
    Black dashed lines, show elliptical fits to the white lines.
    At the highest disk inclinations, the isophotes no longer trace the
    full gap in the polarized intensity and elliptical fits are not made.  
    Eccentricities of the dark and bright isophotes are all
    very nearly equal to $\sin i$.  
  }
\end{figure*}

Rather than relying on true isophotes to interpret disk inclination,
we can use the locations of the gaps as determined from the
brightness minima and maxima as measured radially.
These are not true isophotes because the
brightness is not azimuthally uniform, but it is convenient to refer to
them as such.  These dark and bright isophotes 
are shown as white lines in Figure \ref{fig:gappics}, 
with black dashed lines as the elliptical
fits.  The sets of lines are coincident, indicating that these isophotes
are very well described as ellipses.  The eccentricity of the ellipses
is very well matched by $\sin i$, and are not plotted in Figure
\ref{fig:isoeccen}.  The isophote center offsets scale very
well with $\sin i$ also, as seen in the right panel of
Figure \ref{fig:isooffset}, consistent with an offset solely due to
inclination of a thick disk.  The positions of the bright and dark
isophotes are unaffected by polarization, showing that the existence
and locations of gaps in disks can be deduced from either or
$P$ and $I$ images alone.
Moreover, the offsets for the 70
and 200 $M_{\oplus}$ planets are very well matched, indicating that using
the radial brightness minimum is an excellent way to determine
the position of the gap trough.  The positions of the bright far gap edge
is different between the two planet masses, most likely because the
gap is wider for the higher mass case, thus pushing the location of
the bright isophote further out.  

\begin{figure}
  \centerline{\includegraphics[width=3.3in,trim=0.2in 0in 0.1in 0in]{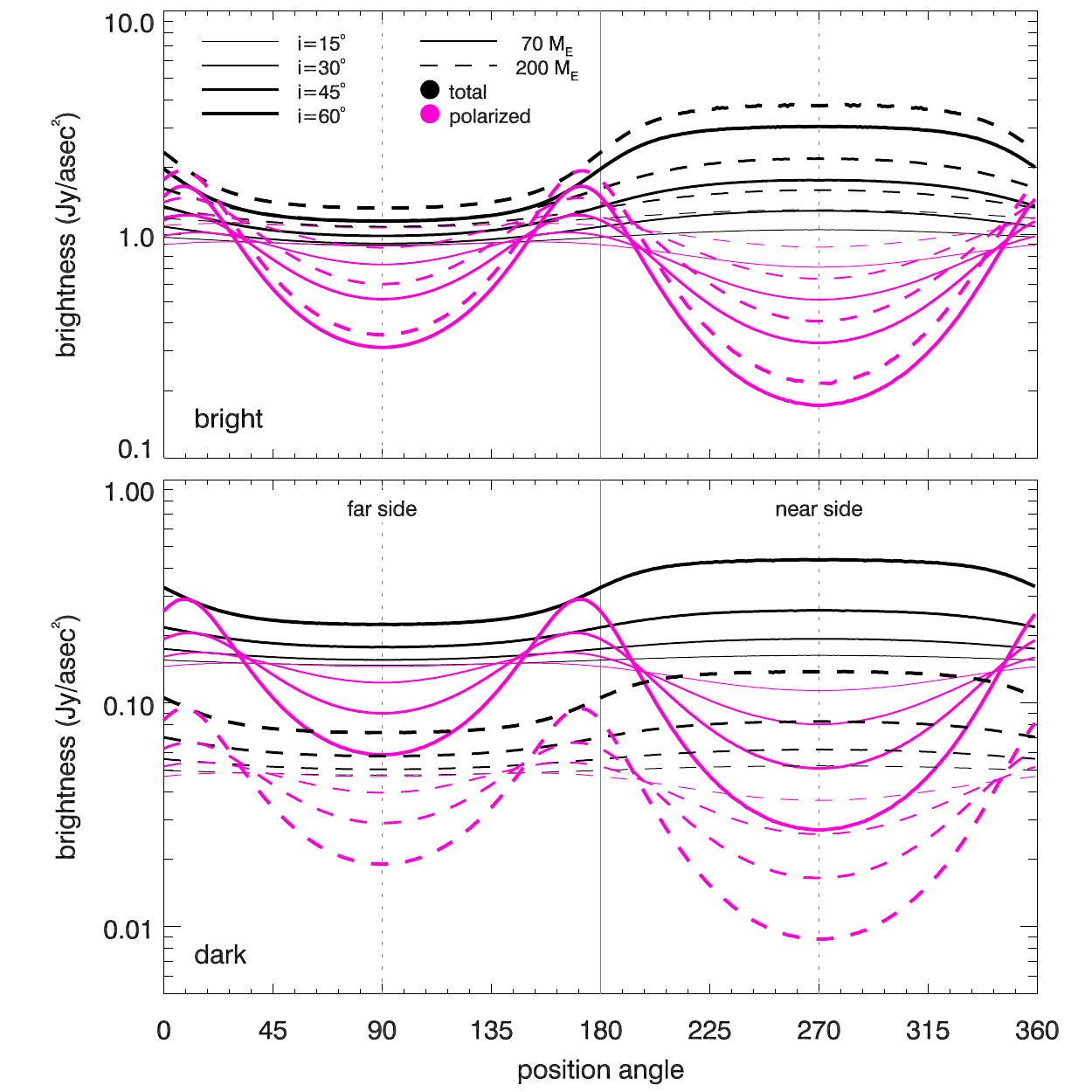}}
  \caption{\label{fig:minmaxbright}
    Azimuthal variation in brightness in the gaps of inclined disks.
    The upper plot traces the brightened far side of the gap and the
    lower plot traces the gap shadow.  Position angle is measured
    from counterclockwise from the right horizontal axis ($+x$) centered
    on the stellar position.  Solid lines and dashed lines
    indicate 70 and 200 $M_{\earth}$ planets, respectively.  Line
    thickness increases with inclination angle.  Black lines show total
    intensity and magenta lines show polarized intensity.  
  }
\end{figure}

To further illustrate how polarization affects images of gapped disks,
we plot the azimuthal brightness profiles along the bright and dark isophotes
in Figure \ref{fig:minmaxbright}.  
In total intensity, the near side of the disk is always brighter than
the far side because the angle of scattering to the observer ($\eta$)
is larger.  This effect was noted in Paper II and can be deduced from
Equation (\ref{eq:totscat}). 
However, the fractional
polarization is lower on the near side of the disk compared
to the far side, which causes the near side to actually appear slightly
dimmer than the far side.
Thus, the $P$ images retain a great deal of symmetry along the
minor axis of the disk image.  This suggests that $P$ images 
alone are not a good method of resolving the near/far side 
degeneracy of disks.  The fractional polarization ($P/I$) is a 
better test of near/far side asymmetry.

Since $P$ images result from a combination of both fractional 
polarization and total intensity, the interpretation 
of an observed $P$ image in the absence of a total intensity 
image may lead to the incorrect interpretation of 
density structure in a disk, when it may, in fact 
be solely a geometric effect.  
On the other hand, polarization information together with 
total intensity imaging may be a valuable 
tool for resolving the near-far degeneracy in inclination angle.

\section{Discussion: Applications to Observations}
\label{sec:disc}

Polarized scattered light images of transitional disks 
can be interpreted in light of the results presented here by
treating the inner clearing as a large gap.  Here, we discuss
a few pertinent examples.  

TW Hya is a well-studied nearly face-on disk that has been imaged
to high resolution in total intensity images \citep{TWHya, TWHya2} and 
polarized intensity \citep{2015Akiyama+, 2016arXiv161008939V}.
Both the total and polarized intensity reveal gaps in the disk at 20 and
80 AU, and the shapes of the radial profiles are roughly consistent
with each other \citep{TWHya2}, as we expect for face-on images.  
\citep{2016Andrews+}.  The gap at 20 AU is also consistent with
ALMA imaging of the disk \citep{2016Andrews+}, suggesting that the
concentric ring structure seen in the disk in scattered light traces
the underlying structure of the disk as a whole.  

The bright ``arcs'' seen in Oph IRS 48 \citep{2015Follette+}
could be the apparent ansae seen along the major axes of the
gapped inclined disks shown in Figures \ref{fig:subcr} and
\ref{fig:mcrit}.  The lack of emission seen along the minor axis
in this disk could be simply due to the low fractional polarization
caused by the inclination of the disk.  

LkCa 15 is an excellent example of how photocenter offsets can be 
caused by finite disk thickness.
The gap imaged in LkCa 15 in scattered polarized light
\citep{2015Thalmann+} traces an ellipse with the same eccentricity as 
seen in 7 mm emission \citep{2014Isella+}, but offset along the minor axis.
This fits with our prediction that the eccentricity of the
bright gap edge matches the underlying shape of the disk, but offset
because of the disk thickness.
In this interpretation, the northern edge of the disk is tilted
toward the observer, and assuming an inclination of 45\degr
and an apparent offset of 11.1 AU in projection, the height
of the disk at the gap edge is 16 AU\@.  If the radius of the
inner hole is 50 AU, this gives $h/r=0.32$.  Given that the height of
the light scattering surface ($h$) can be several times the thermal scale
height of the disk, this is not an unreasonable finding.  

PDS 66 shows a strong brightness asymmetry along the minor axis
of the ring seen at $\sim70-90$ AU (in projection) 
in total intensity images \citep{2014Schneider+}.  However, this
asymmetry is less strong in polarized intensity \citep{2016Wolff+}.  
The brighter eastern edge (in total intensity) could be the side tilted
away from the observer, resulting in a smaller polarization fraction
on that side than the other.  This would also explain the positional
offset of the inner bright arc seen at $\sim20$ AU\@.  
The eastern edge still appears to be the azimuthal brightness maximum in 
the polarized image, which could be due to some kind of highly
back-scattering grains or real anisotropies in the disk.  Given other
azimuthal anisotropies seen in the disk, a true structural anisotropy
could be the best explanation.

\section{Conclusions}
\label{sec:concl}

Scattered polarized light is a promising way to detect and 
characterize protoplanetary disks because of the increased contrast, 
but interpreting these images for structure must be undertaken with 
caution.  The changing scattering angle and polarization 
fraction across the disk can introduce spurious structure in the disk.  
If polarized light is the only method used to image
a protoplanetary disk, then any observed morphology must be treated very
carefully before inferring any actual disk structure.

Face-on disks with gaps have show slightly
less contrast across the gap in polarized intensity
than in scattered intensity because the brightened far edge
is less polarizing than elsewhere in the disk.  This is primarily
due to the fact that light scattered from the brightened far edge
has a higher fraction of multiply scattered photons.  
  
Inclined disks without gaps appear to be more eccentric when viewed
in polarized intensity as opposed to total intensity, as measured by
the eccentricity of isophotes.  Measuring the eccentricity of isophotes
in {\em total intensity} 
robustly recovers the inclination of the disk.  However, the thickness 
of the disk results in a photocenter offset from the central star.
Any photocenter offset measured along the minor axis of a disk image
can therefore be interpreted as a measure of disk thickness rather
than an indication of intrinsic eccentricity.
Isophotes in polarized intensity images are more eccentric than
those measured in total intensity.  At low inclination angles, the
photocenter offset is larger when viewed in polarized intensity.
At high inclinations, isophotes are no longer well-described as ellipses.
This can create the appearance of apparent structure in the disk
when none is truly there.  

Partially cleared gaps in inclined disks,
such as those created by sub-Jovian-mass planets,
can give the appearance of
local holes or voids of emission where
no such structural features exist. 
On the other hand, since the apparent position of a gap,
as measured by the radial brightness minima and maxima, is
unaffected by polarization, the presence of a gap can be inferred
in either total or polarized intensity, so long as the entire
gap can be resolved.  In total intensity images, the near side
of the disk will appear brighter than the far side
within the gap.  However, in polarized
light, the brightness of the gap will appear more symmetric because the
near side of the gap scattered less polarized than the far side.

In principle, if both the polarized and total intensity of a disk image
are known along with the scattering angles deduced from the disk inclination,
then the scattering phase function can also be determined.  This would
constrain the properties of the dust itself, leading to a better
understanding of the composition of protoplanetary disks.  
Thus, polarized scattered light can be a useful probe of the
dust properties in planet-forming disks.

When interpreting disk structure from polarized scattered light images
alone, one must account for the effects of disk inclination and 
changing fractional polarization across the surface before
deducing structure in the disk.
When these factors are properly accounted for, imaging of disks
in polarized light can become a powerful tool for understanding
protoplanetary disks.  

\acknowledgements
The author thanks John Debes, for helpful comments in the preparation
of this paper, and an anonymous referee, whose feedback greatly
improved this paper.  
This work was supported by NASA XRP Grant NNX15AE23G.

\bibliographystyle{aasjournal}
\bibliography{../../planets,../../jang-condell,pol}

\end{document}